\documentclass[aps,prl,twocolumn,groupadress,showpacs]{revtex4-1}
\usepackage{setspace}
\usepackage{graphicx}
\usepackage{epstopdf}
\usepackage{amsmath}
\usepackage{amssymb}
\usepackage{epsfig}
\usepackage{dcolumn}
\usepackage{amsmath}
\usepackage{graphicx}

\usepackage{color}
\usepackage[normalem]{ulem}

\begin{document}

\title{Inversion-symmetry-breaking-activated shear Raman bands in
$T'$-MoTe$_2$}

\author{Shao-Yu Chen$^{1}$}
\author{Thomas Goldstein$^{1}$}
\author{Ashwin Ramasubramaniam$^{2}$}
\author{Jun Yan$^{1}$}
\email{yan@physics.umass.edu}

\affiliation{$^1$ Department of Physics, University of Massachusetts, Amherst, MA 01003, USA  \\
$^2$ Department of Mechanical \& Industrial Engineering, University of Massachusetts, Amherst, MA 01003, USA
}

\date{\today}

\begin{abstract}
Type-II Weyl fermion nodes, located at the touching points between
electron and hole pockets, have been recently predicted to occur in
distorted octahedral ($T'$) transition metal dichalcogenide
semimetals, contingent upon the condition that the layered crystal
has the noncentrosymmetric orthorhombic ($T'_{or}$) stacking. Here,
we report on the emergence of two shear Raman bands activated by
inversion symmetry breaking in $T'$-MoTe$_2$ due to sample cooling.
Polarization and crystal orientation resolved measurements further
point to a phase transition from the monoclinic ($T'_{mo}$)
structure to the desired $T'_{or}$ lattice. These results provide
spectroscopic evidence that low-temperature $T'$-MoTe$_2$ is
suitable for probing type-II Weyl physics.
\end{abstract}


\maketitle

\newcommand{\incl}{\includegraphics}

$T'$ transition-metal dichalocgenides ($T'$-TMDCs), in which the
metal atoms are octahedrally coordinated and subject to a
Peierls-like metal-metal zigzag bonding distortion, are emerging as
a class of materials with intriguing non-trivial band topology.
Monolayer $T'$-TMDCs have been shown theoretically to be two
dimensional topological insulators \cite{Qian2014} whereas bulk
$T'$-TMDCs with appropriate layer stacking are predicted to be
type-II Weyl semimetals \cite{Soluyanov2015}. Originally a
fundamental high-energy-physics concept in quantum field theory
\cite{Weyl1929}, Weyl fermions have recently been proposed and
demonstrated as a new type of condensed matter quasi-particle
excitation \cite{Weng2015, Huang2015, Xu2015,
Lv2015,Yang2015,Lv2015a, Xu2015a, Xu2015b}. In these solid state
systems, paired Weyl nodes are located at distinct positions in the
crystal momentum space with each node acting as a `magnetic
monopole' from which Berry flux emanates in directions determined by
the node's chirality. This leads to anomalous phenomena including
Fermi arcs on separate crystal surfaces connected through the bulk
\cite{Wan2011}, and apparent violation of charge conservation for
electrons with a definite chirality propagating in parallel electric
and magnetic fields \cite{Nielsen1983}.

Depending upon the material-specific Fermi velocity tensor, the Weyl
nodes can either appear as a point-like Fermi surface (type-I) or as
the touching points between electron and hole pockets
(type-II)\cite{Soluyanov2015}. Type-I Weyl nodes have been
demonstrated in recent experimental studies of TaAs \cite{Xu2015,
Lv2015,Yang2015,Lv2015a}, NbAs \cite{Xu2015a} and TaP
\cite{Xu2015b}, while type-II Weyl nodes, a more recent theoretical
development \cite{Soluyanov2015}, remains to be confirmed
experimentally. Distorted octahedral orthorhombic ($T'_{or}$)
WTe$_2$ was the first material predicted to be a type-II Weyl
semimetal \cite{Soluyanov2015}. Subsequent theoretical
investigations showed that the $T'_{or}$ phase of MoTe$_2$ also
possess type-II Weyl nodes \cite{Sun2015}. Numerical calculations
reveal that $T'_{or}$-MoTe$_2$ has much larger Weyl-node-pair
separations than WTe$_2$ \cite{Soluyanov2015,Sun2015}.
Closely-spaced Weyl nodes may become more vulnerable to internode
scattering, and lead to short Fermi arcs, making it more challenging
to investigate with tools such as angle-resolved photo emission
spectroscopy. This makes $T'_{or}$-MoTe$_2$ potentially a more
promising candidate for investigating the quantum behavior of Weyl
fermions.

However, unlike WTe$_2$ that has the desired $T'_{or}$ phase for
Weyl nodes, semimetallic MoTe$_2$ crystallizes in the undesirable
monoclinic ($T'_{mo}$) phase (also called $\beta$-MoTe$_2$) at room
temperature \cite{Brown1966}; a recent study further indicates that
this $T'_{mo}$ phase prevails in $T'$-MoTe$_2$ down to 1.8 Kelvin
\cite{Keum2015}, while earlier investigations suggested that the low
temperature structure might be altogether different \cite{
Hughes1978, Clarke1978}. The $T'_{mo}$ lattice has both
time-reversal symmetry and inversion symmetry, making it impossible
to have the topologically non-trivial electronic bands for a Weyl
semimetal. This undesired $T'_{mo}$ phase, in fact, prompted the
investigation of Mo$_x$W$_{1-x}$Te$_2$ that inherits the $T'_{or}$
structure of WTe$_2$ and enables large Weyl-node-pair separations
\cite{Chang2015,Belopolski2015}. From an experimental point of view,
MoTe$_2$ is a stoichiometric compound and would be preferred over
Mo$_x$W$_{1-x}$Te$_2$ for investigating Weyl physics provided that
its $T'_{or}$ phase can be established rigorously. Presently, the
existence of orthorhombic $T'$-MoTe$_2$ is being debated
\cite{Chang2015,Belopolski2015,Sun2015,Wang2015} and more
experimental studies are needed to elucidate the low-temperature
phase of the crystal.

In this Letter, we make use of Raman scattering to systematically
monitor the inversion symmetry and the crystal phase of
$T'$-MoTe$_2$ as a function of temperature.  At low temperatures,
two low-energy modes (at 12.6 and 29.1\,cm$^{-1}$ respectively)
associated with interlayer shear vibrations, one perpendicular and
the other parallel to the distorted zigzag metal chains, appear in
the spectra. We conclude from symmetry analysis that the appearance
of the two Raman bands is a consequence of inversion symmetry
breaking. The thermal cycling as well as the polarization-resolved
and crystal-orientation-resolved Raman tensor analysis further
confirm that during cooling, a structural phase transtion from
$T'_{mo}$ to $T'_{or}$ has occurred. These findings provide strong
evidence for the low temperature Weyl $T'_{or}$ phase of the
semimetallic MoTe$_2$. Furthermore, our experiments offer insights
into inversion symmetry breaking in $T'$-MoTe$_2$ that can be of
significance to nonlinear optics \cite{Kumar2013,Li2013}, and open
up a venue for investigating stacking dependent vibrational, optical
and electronic properties of $T'$-MoTe$_2$ atomic layers.

\begin{figure}
\centering \includegraphics[scale=0.4]{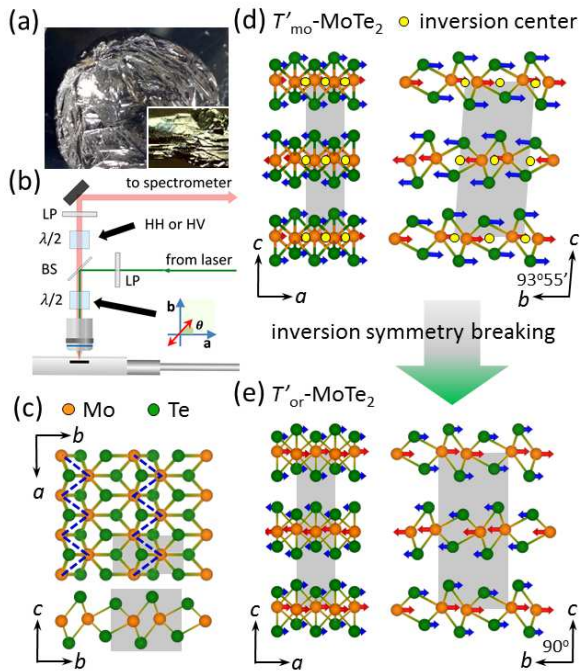}\caption{ (a)
Pictures of bulk $T'$-MoTe$_2$. Inset is a zoom-in for showing
detailed sample morphology. (b) Schematic of polarization-resolved
and crystal-orientation-resolved Raman spectroscopy. (c) The top
view ($a$-$b$ plane) and side view ($b$-$c$ plane) of monolayer
$T'$-MoTe$_2$. The blue dashed lines indicate the Mo-Mo zigzag chain
which is aligned with the $a$-axis. (d)\&(e) The shear vibrations in
$T'_{mo}$ (d) and $T'_{or}$ (e) MoTe$_2$. (left: shear mode along
$a$-axis; right: shear mode along $b$-axis). The shaded areas in
(c)-(e) show the unit cell of $T'$-MoTe$_2$ in the corresponding
phases. The inversion symmetry is broken in $T'_{or}$-MoTe$_2$.}
\label{fig1}
\end{figure}

The $T'$-MoTe$_2$ crystal used in this work is grown by chemical
vapor transport method using Bromine as the transport agent. At room
temperature the most stable phase of MoTe$_2$ is the hexagonal
semiconducting phase (also called $\alpha$-MoTe$_2$). To crystalize
in the metastable semimetallic phase we thermally quench the crystal
from 900\,$^{\circ}$C to room temperature with water bath. The image
of a typical crystal is shown in Fig.1(a). The crystal shows layered
structure and needle-like shape with lengths ranging from a few mm
to 1\,cm.

Figure 1(b) schematically shows our experimental setup. The
$T'$-MoTe$_2$ crystals are mounted in a microscopy cryostat with the
flat plane facing up. Linearly polarized light from a frequency
doubled Nd:YAG laser at 532\,nm is used to excited the sample in
back-scattering geometry. We place a half waveplate between the beam
splitter (BS) and the microscope objective to adjust the angle
$\theta$ of the incident light polarization direction with respect
to the crystal $a$-axis. Another half waveplate is mounted in the
collection path after the beam splitter followed by a linear
polarizer (LP) so that we can selectively collect the scattered
light that is polarized either parallel (HH) or perpendicular (HV)
to the incident beam. The scattered light is then dispersed by
Horiba T64000 triple spectrometer operating in subtractive mode, and
detected by a liquid nitrogen cooled CCD camera.

Figure 1(c) shows the top and side views of the atomic arrangements
in a $T'$-MoTe$_2$ monolayer. The in-plane $a$ and $b$ axes are
parallel and perpendicular to the zigzag Mo-Mo chains respectively.
In bulk crystals, the neighboring MoTe$_2$ layers are rotated from
one another by 180$^{\circ}$ about an axis perpendicular to the
atomic layer.  This results in two MoTe$_2$ layers per bulk unit
cell, as enclosed by the shaded parallelogram and rectangles in
Figs.1 (d)\&(e). Consequently both $T'_{mo}$ and $T'_{or}$ structures
support shear mode vibrations as illustrated by the arrows in the
drawings. We note that due to in-plane anisotropy in $T'$ crystals,
the shear mode vibrations along the $a$ and $b$ axes are
non-degenerate, in contrast to the hexagonal phase
\cite{Zhao2013,Chen2015,Froehlicher2015}.

The $T'_{mo}$ and $T'_{or}$ phases differ in their $c$-axis
directions: in the former the $c$-axis makes an angle of
93$^{\circ}$55$'$ with the $b$-axis, whereas in the latter the
$c$-axis is perpendicular to the atomic plane. While differing only
by a small shift of the atomic layers along the $b$-axis direction,
the two phases have important differences in symmetry. In
particular, $T'_{mo}$ is inversion symmetric (inversion centers are
noted as yellow dots in Fig.1(d)) while $T'_{or}$ is
non-centrosymmetric. This inversion symmetry is the key reason why
$T'_{mo}$-MoTe$_2$ can not be a Weyl semimetal \cite{Sun2015}, and
the breaking of this symmetry is a focus of this paper.

\begin{figure}
\centering \includegraphics[scale=0.42]{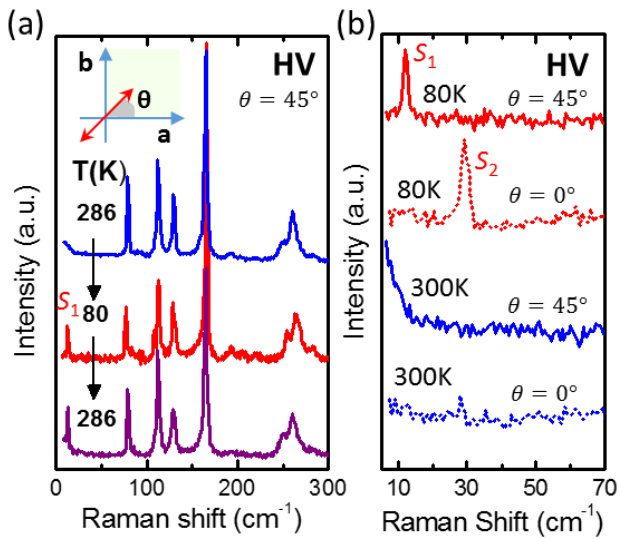} \caption{  (a) The
Raman spectra of $T'$-MoTe$_2$ under different thermal cycles. The
low wavenumber shear mode ($S_1$) emerges when cooling down from
286\,K (blue) to 80\,K (red) and persists during warming up to
286\,K (purple). (b) The Raman spectra of low wavenumber modes of
$T'$-MoTe$_2$ at 80\,K and 300\,K at different crystal orientation
angles. The spectra in both figures are taken in the HV
configuration.
 } \label{fig2}
\end{figure}

Figure 2(a) shows typical Raman spectra of $T'$-MoTe$_2$, collected
as the crystal is cooled down from 286\,K to 80\,K and warmed back
up to 286\,K, with incident and scattered light linearly polarized
perpendicular to each other (HV), and making a 45$^{\circ}$ angle
with the $a$-axis of the crystal. Because of the relatively large
number of atoms (12) in the $T'$ unit cell as compared to the
hexagonal phase, the spectra display many first order phonon bands
below 300\,cm$^{-1}$. In this paper we will focus on the low
wavenumber regime (Fig.2(b)) wherein the symmetry-breaking between
the $T'_{or}$ and $T'_{mo}$ phases is readily discerned; systematic
analysis of the high-energy optical phonons will be presented
elsewhere. As seen in Fig. 2(a), cooling induces a new Raman peak
$S_1$ at 12.6\,cm$^{-1}$ in the spectrum and it persists when the
sample is warmed back up. In Fig.2(b), we also display spectra with
the incident and scattered light linearly polarized along the $a$
and $b$ axes respectively (HV, $\theta$=0$^{\circ}$). In this
detection configuration, we observe another new Raman band $S_2$ at
29.1\,cm$^{-1}$ at low temperature.

To understand the origin of these new vibrational features, we
performed lattice dynamics calculations using plane-wave density
functional theory (DFT) as implemented in the Vienna Ab Initio
Simulation Package (VASP)\cite{Kresse1996,footnote}. As standard DFT
functionals fail to describe interlayer van der Waals bonding
correctly, we used the non-local optB86b van der Waals
functional\cite{Klimes2011,Dion2004}, which reproduces the
equilibrium geometry of MoTe$_{2}$ accurately\cite{Bjorkman2014}. By
scrutinizing the normal modes from our DFT calculations, we find
that in the $T'_{mo}$ phase the shear modes along the $b$ and $a$
axes are at 9.2 \,cm$^{-1}$ and 27.2 \,cm$^{-1}$, respectively,
whereas in $T'_{or}$ phase the two modes are found at 15.3 and 29.3
\,cm$^{-1}$. Therefore, we attribute the experimentally observed
low-energy $S_1$ and $S_2$ peaks in Fig.2 to the interlayer shear
modes along the $b$ and $a$ axes respectively.

Because both $T'_{mo}$ and $T'_{or}$ phases host these shear
vibrations as illustrated in Fig.1(d)\&(e) and confirmed by the DFT
calculation, it is not obvious to which crystal phase the Raman
peaks in Fig.2 belong. Raman intensity of materials can have unusual
temperature dependencies \cite{Lui2014}. Could it be that the
crystal is in the same $T'_{mo}$ phase at all temperatures, and that
its shear modes somehow only have observable Raman intensity at low
temperature? The data in Fig.2(a) suggest that this is not the case:
$S_1$ is absent in the original warm sample at 286\,K, shows up at
80\,K, and persists as the sample is warmed back up to 286\,K. The
significant hysteresis (more details in Fig.4) indicates that the
appearance of $S_1$ and $S_2$ are due to structural, instead of
thermal reasons, i.e., at the same temperature the shear modes may
or may not appear depending on the underlying crystal structure.

The reason why $S_1$ and $S_2$ do not appear in $T'_{mo}$-MoTe$_2$
Raman spectra can be understood from symmetry. The inversion centers
of the monoclinic structure (Fig.1(d) yellow dots) are located
within the MoTe$_2$ layer. As a result, $S_1$ and $S_2$ vibrations
change their directions under inversion. These odd shear modes are
thus infrared (IR) active. IR and Raman active modes in
centrosymmetric crystals are mutually exclusive, which explains the
absence of $S_1$ and $S_2$ in Raman spectra of $T'_{mo}$-MoTe$_2$ in
Fig.2. This is in contrast to, say, hexagonal bilayer TMDCs, whose
inversion centers are located between the TMDC layers, and which,
consequently, have even shear modes that are Raman
active\cite{Zhao2013, Chen2015, Froehlicher2015}.
The appearance of $S_1$ and $S_2$ Raman bands due to sample cooling
thus presents a signature for inversion symmetry breaking, which
incidentally and importantly, is also a necessary (albeit not
sufficient) condition for the existence of Weyl fermions in a
non-magnetic material.

To further establish that the two low-energy Raman peaks are indeed
linked to the Weyl $T'_{or}$-MoTe$_2$ phase, we perform polarization
and crystal-orientation resolved Raman tensor analysis. The
$T'_{or}$-TMDC structure has three symmetry operations including a
mirror plane ($m$), a glide plane ($n$) and a two-fold screw axis
($2_1$) that form the $C_{2v}^7$ group (No. 31 $Pmn2_1$ space
group)\cite{Brown1966}. Performing these operations explicitly upon
the atomic displacements in Figs.1(d)\&(e), we found that $S_1$
along the $b$ axis is symmetric under all three operations, while
$S_2$ along the $a$ axis is symmetric under $2_1$ and antisymmetric
under the other two operations. Hence $S_1$ is expected to have
$A_1$ symmetry and $S_2$ should be $A_2$.

The Raman tensor $\cal{R}$ for the $A_1$ and $A_2$ modes of $C_{2v}$ point group are given
respectively by \cite{Loudon1964}:
\begin{eqnarray} \label{eq:RA1A2}
{\cal{R}}(A_1)=\left(\begin{array}{ccc} d & 0 &0 \\ 0&e&0 \\
0&0&f \end{array} \right), {\cal{R}}(A_2)=\left(\begin{array}{ccc} 0 & g &0 \\ g&0&0 \\
0&0&0 \end{array} \right).
\end{eqnarray}
The Raman cross-section is expressed as \linebreak ${\cal
I}=|<\epsilon_i|{\cal R}|\epsilon_o>|^2$, where $\epsilon_i$ and
$\epsilon_o$ are polarization states of the incident and scattered
light. Consider backscattering geometry using linearly polarized
light with $\epsilon_i=(\cos\theta, \sin\theta, 0)$ and
$\epsilon_o=(\cos\phi, \sin\phi, 0)$, where $\theta$ and $\phi$ are
angles between the direction of MoTe$_2$ crystal $a$-axis and that
of the light polarization. The Raman cross sections for $A_1$ and
$A_2$ are:
\begin{eqnarray} \label{eq:IA1A2}
{\cal{I}}(A_1)=\left(d\cos\theta\cos\phi+e\sin\theta\sin\phi
\right)^2, \nonumber \\
{\cal{I}}(A_2)=\left(g\sin(\theta+\phi)\right)^2.
\end{eqnarray}
%
\begin{figure}{
\centering \includegraphics[scale=0.4]{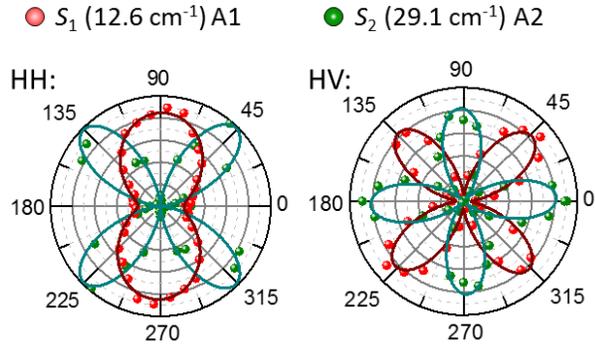}} \caption{
Polarization (HH or HV configuration) and crystal orientation
dependent intensity of shear modes, $S_1$ (red) and $S_2$ (green).
The solid curves are fits using equation (3) in the text. The
numbers on the angular axis indicate the angle relative to the
$a$-axis of crystal, as denoted by $\theta$ in Fig.2. } \label{fig3}
\end{figure}
%
In our measurement, HH configuration corresponds to $\phi=\theta$,
and HV corresponds to $\phi=\theta+\frac{\pi}{2}$ in Eq.(2). This
gives
\begin{eqnarray} \label{eq:ISaSb}
{\cal{I}}_{HH}(S_1)=\left(d\cos^2\theta+e\sin^2\theta
\right)^2, \nonumber  \\
{\cal{I}}_{HV}(S_1)=\left(\frac{d-e}{2}\right)^2\sin^2(2\theta),\nonumber
 \\ {\cal{I}}_{HH}(S_2)=g^2\sin^2(2\theta) , \nonumber \\
{\cal{I}}_{HV}(S_2)=g^2\cos^2(2\theta).
\end{eqnarray}
Figure 3 presents the angular dependence of the Raman intensities as
well as the fits according to Equation (3). In HH configuration, the
$S_1$ mode shows two-fold symmetry while $S_2$ mode shows four-fold
symmetry. In contrast, both $S_1$ and $S_2$ exhibit four-fold
symmetry albeit with 45$^{\circ}$ shift in HV scattering geometry.
The excellent agreement between data and theoretical calculation
verifies the symmetry properties of the two shear modes, and
confirms that the observed $S_1$ and $S_2$ Raman bands are from the
$T'_{or}$ phase of the MoTe$_2$ crystal.

\begin{figure}{
\centering \includegraphics[scale=0.4]{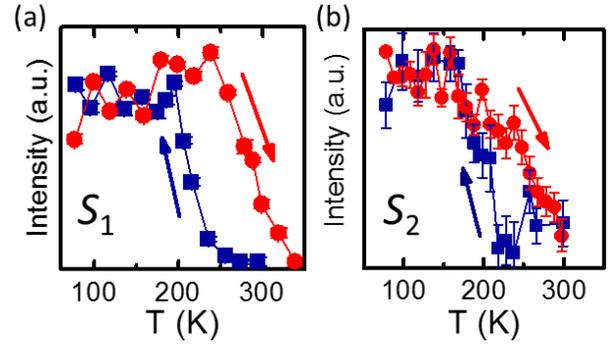}} \caption{
Temperature dependent intensity of shear modes. (a)$S_1$ and
(b)$S_2$ under different thermal cycle, cooling (blue) and warming
(red). The hysteresis means that both $T'_{mo}$ and $T'_{or}$ phases
can coexist in certain temperature range. For temperatures lower
than 150\,K, the intensity overlaps for cooling and heating,
indicating a complete phase transition from $T'_{mo}$ to $T'_{or}$.}
\label{fig4}
\end{figure}

To gain more insight into the inversion-symmetry-breaking phase
transition from $T'_{mo}$ to $T'_{or}$, we monitor the $S_1$ and
$S_2$ peaks when the sample is cooled down from room temperature and
warmed back up. Figure 4 shows the temperature dependence of the
$S_1$ and $S_2$ intensities as a function of temperature. As can
been seen, the cooling (blue) and warming (red) curves do not
overlap over extended range of temperatures. The thermal hysteresis
suggests that the $T'_{mo}$ to $T'_{or}$ and $T'_{or}$ to $T'_{mo}$
phase transitions take time to complete and the crystal lattice
cannot adjust very quickly from one stacking to another. The
intensities of the Raman bands tend to coalesce below 200\,K and,
moreover, the intensity below 150\,K is independent of cooling or
warming. This indicates that the crystal is stabilized in the pure
$T'_{or}$ phase, without any admixtures from the $T'_{mo}$, which is
important for obtaining high quality $T'_{or}$-MoTe$_2$ crystal to
probe the Weyl physics.

In conclusion, the inversion symmetry and the crystal phase of
$T'$-MoTe$_2$ was probed by Raman scattering. The two new shear
modes that we observed and systematically analyzed provide strong
evidence for the emergence of the orthorhombic $T'$-MoTe$_2$ phase
upon cooling of the room-temperature monoclinic phase. This
investigation opens up promising opportunities to investigate the
theoretically predicted and yet to be experimentally confirmed
type-II Weyl semimetal. We further anticipate that the
cooling-driven inversion-symmetry breaking might also be probed by
second harmonic generation \cite{Kumar2013,Li2013}. Finally, the
thermally-driven stacking changes could also occur in
atomically-thin $T'$-MoTe$_2$, raising interesting questions
regarding stacking-dependent vibrational, optical and electronic
properties, which are know to display rich physics in, for example,
another 2D semimetal graphene \cite{Zhang2011, Lui2011, Bao2011}.

{\small \textbf{Acknowledgements}

This work is supported by the University of Massachusetts Amherst
and in part by the National Science Foundation Center for
Hierarchical Manufacturing (CMMI-1025020). Computing support from the
Massachusetts Green High Performance Computing Center is gratefully acknowledged.
}

\bibliography{mote2weylAR1b}

\end{document}